\documentstyle[preprint,aps]{revtex}
\def\x{{\bf x}}
\def\v{\tilde{\bf v}}
\begin{document}
\draft
\preprint{titcmt-94-*}
\tighten
\title{Power law velocity fluctuations due to inelastic collisions
 in numerically simulated vibrated bed of powder}
\author{Y-h. Taguchi\cite{e-mail}}
\address{Department of Physics,
Tokyo Institute of Technology\\
Oh-okayama, Meguro-ku, Tokyo 152, Japan}
\date{\today}
\author{Hideki Takayasu}
\address{Graduate School of Information Sciences\\
Tohoku University, Sendai 980-77, Japan}
\maketitle
\nopagebreak
\begin{abstract}
Distribution functions of relative velocities among
particles in a vibrated bed of powder are studied both
numerically and theoretically. In the solid phase where
granular particles remain near their local stable states,
the probability distribution is Gaussian. On the other hand,
in the fluidized phase, where the particles can exchange
their positions, the distribution clearly deviates from Gaussian.
This is interpreted with two analogies;
aggregation processes
and soft-to-hard turbulence transition in thermal convection.
The non-Gaussian distribution is well-approximated
 by the t-distribution which is derived theoretically
 by considering the effect of clustering by inelastic collisions
in the former analogy.
\end{abstract}
\narrowtext

Granular matter are attracting  much interest of
physicists\cite{review_powder}.
Among them, people extensively study vibrated bed of powder,
which
consists of a vessel filled with  granular
matter and a loud speaker to shake it.
When the acceleration amplitude of the
vibration $\Gamma$, the control parameter, exceeds
some critical value,
the bed is fluidized and can show  many interesting phenomena:
 heaping,  convection\cite{exp_bed},
 capillarity\cite{Akiyama},
 surface fluidization\cite{Evesque.epl,Luding},
 Brazil nuts segregation\cite{Rosato,Duran,Jullien,Ohtsuki,Knight},
 standing waves\cite{Douady,Pak,Swinney,Hubler}, and so on.

In this letter we first report that
dynamical phase transition takes place in numerically simulated
vibrated bed of powder: probability distribution function (PDF) of
displacement vector changes from Gaussian to non-Gaussian.
This phase transition is interpreted in two ways:
a soft-to-hard turbulence transition
similar to that of fluid turbulence
and the averaging effect of particle
velocities due to inelastic clustering of  granules.

The numerical setup (2D) used in
this letter has been described
in detail in Refs. \cite{tag_turbulence}.
Numerical simulations  not  only reproduce
convection\cite{tag_prl,Gallas},
but also show powder
turbulence\cite{tag_turbulence,tag_JdP}.
The particles interact by
 visco-elastic collisions\cite{model},
and their motions are integrated
under the existence of  gravity acceleration $g$.
The material parameters are
the collision time $t_{col}$,
the time period during collision,
and the coefficient of restitution $e$.
The bed is two-dimensional, with
a periodic condition along the horizontal
direction.
The bottom of the vessel oscillates
vertically as a function
of time $t, b \cos \omega_0 t$,
where $\Gamma = b \omega_0^2$.

      In order to keep all of the particles in a fixed area,
a lid ( or a weight) is put horizontally on the granular layer.
The spacing between the lid and
the bottom fluctuates as particles hit the lid.
This spacing can be used as a guide
to observe stationary statistics; namely,
we observe the locations of the particles at
a time when the spacing takes a value
in a given small interval $[h_0, h_0 + dh_0]$.
Let the $n$th observation time be $t(n)$ and
the $i$th particle's location at time $t(n)$ be $\x_i(n)$,
then the relative velocity $\v_i(n)$ is defined
by the displacement   $\Delta \x_i(n) = \x_i(n) - \x_i(n-1)$
over the time interval $\Delta t(n) = t(n) - t(n-1)$.

     Our numerical simulation is performed with
the following parameters: $g=1.0, \omega_0 = 2\pi/6, b = 1.0 (\Gamma =1.1)$,
the total number of particles $N_{tot} = 1024$,
the horizontal width of the bed $L_h =128$,
the particle diameter $d = 2.0$,
the particle mass $m = 1, t_{col} = 0.1, e=0.8$,
the mass of the lid $M = 100$, and $h_0 = 30.0$.
With these parameters we can find two phases coexisting in the vessel:
the lower region of the layer
belongs to the solid phase in which no pair of
particles exchange their position,
and the upper region belongs to the fluidized phase
where particles exchange their positions and show fluid-like
 collective motions.
These two phases both have
the Kolmogorov's power spectrum, $k^{-5/3}$\cite{tag_turbulence},
where $k$ denotes the wave number for the relative velocity
field  $\v_i^{(n)}$.
Both phases are thus regarded as turbulent states.
However, they have
completely different PDFs ;
the PDF in the solid region
is close to Gaussian,
but the fluidized region has a
PDF very different from  Gaussian(Fig. \ref{fig:PDF_1}).
For larger $\Gamma$,
($\Gamma=1.64$: $b$ is changed to be $1.5$ from $b=1.0$
when $\Gamma=1.1$)
the bed is now fully fluidized and
we can find non-Gaussian PDFs (Fig. \ref{fig:PDF_2}).

In order to understand this phenomenon, we try to interpret it in
two ways. First analogy is aggregation processes of inelastically
colliding particles.
Actually speaking, it  recently turns out
that inelastically colliding particles
can exhibit non-Gaussian PDF of velocity\cite{non-Gauss.inelastic}.
Thus this analogy seems to be suitable for understanding this phenomenon.
In order to explain the non-Gaussian
tails theoretically,
let us assume that $N$ particles having velocities
$\{v_i\}$ exchange
their momenta frequently through mutual inelastic collisions.
By the conservation of total momenta
it is obvious that each velocity $v_i$ gets
closer to the averaged velocity,
$V=1/N \sum v_i$, as inelastic collisions are repeated.
Assuming also that the velocity distributions
for   ${v_i}$ before collisions follow  independent
identical Gaussian with the mean value $v=0$ and variance $\sigma^2$.
Then the distribution of the averaged velocity $V$ is given as
\begin{equation}
P_N(V) = \frac{1}{\sqrt{2 \pi \sigma^2/N}}
\exp\left(-\frac{V^2}{2 \sigma^2/N}\right).
\label{eq:P(V)}
\end{equation}
For a fixed value of  $N$  this PDF is also Gaussian,
however, the value of $N$, which can be regarded as
the size of a cluster,%\cite{cluster}
 should be a random number.
As a kind of mean-field approximation we assume that
any pair of particles belong to the same cluster
with an independent and identical probability.
With this assumption the probability that
a particle is a component of cluster of size $N$, $W(N)$,
is given by an exponential function as, $W(N)=c \; \exp(-cN)$,
where $c$ is a positive constant.
Then, the probability of finding a particle
with velocity $v$ after collisions is given as
\begin{equation}
P(v)=\sum_N  W(N) \frac{1}{\sqrt{2 \pi \sigma^2/N}}
\exp\left(-\frac{v^2}{2 \sigma^2/N}\right).
\label{eq:PDF_v}
\end{equation}
Approximating the summation in Eq.(\ref{eq:PDF_v})
by the integral   $\int dN$  we get the following functional form:
\begin{equation}
P(v)=\frac{1}{2a} \left[ 1 +\left(\frac{v}{a}\right)^2\right]^{-3/2},
\label{eq:P(V).result}
\end{equation}
where $a = \sigma \sqrt{2c}$. Note that this distribution
is close to Gaussian in the vicinity of $v=0$, however,
for large $|v|$ it has power law tails of  $|v|^{-3}$.
In Fig. \ref{fig:PDF_2} this theoretical estimate of PDF\cite{norm} is compared
with the PDF in the hard turbulent phase of our numerical experiments.
We can find a good fit in the whole range of  $v$,
which validates our clustering assumption.

{}From a mathematical view point the PDF in Eq.
(\ref{eq:PDF_v})
is a special case of a t-distribution.
A general t-distribution can be derived
by considering the following generalized cluster  distribution as
\begin{equation}
W(N) \sim N^{\alpha-1} \exp(-cN).
\label{eq:gen_W(N)}
\end{equation}
For $1 \geq \alpha > 0$ the distribution is a decreasing function
like the exponential distribution ($\alpha =1$);
however, for $\alpha > 1$ the distribution has
a maximum around $\alpha/c$.
The particle velocity distribution with
the general cluster distribution Eq.(\ref{eq:gen_W(N)}) becomes
\begin{equation}
P(v) \sim [ 1+(v/a)^2]^{-\alpha -1/2}
%\label{eq:gen_PDF}
\end{equation}
This is the general form of the t-distribution.
This PDF converges to a Gaussian
in the limit of  $\alpha \rightarrow \infty$, which corresponds to
an infinite cluster size.
For any finite $\alpha$,  $p(v)$ is close to Gaussian around  $v=0$,
however, there always exist long tails in the power law.

The second analogy is thermal convection turbulence.
Thermal convective turbulence  is
categorized into two phases according to their PDF\cite{soft-to-hard}.
In cases where the temperature difference between
the upper and the lower boundaries of a fluid container is
intermediate, we can find the so-called soft turbulent phase,
in which  thermal convective flows show irregular fluctuations
following Gaussian distributions.
When the temperature difference is enlarged, another phase,
the so-called hard turbulent phase,
appears; in this phase the fluctuation clearly deviates
from Gaussian showing long tails.
Although much effort has been made to understand this phenomenon,
the underlying mechanism of producing
the non-Gaussian fluctuations has yet to be elucidated\cite{review}.

In our granular bed, it has already numerically been confirmed that
both solid and fluid phase exhibits Kolmogorov's -5/3 powder
spectrum\cite{tag_turbulence}.
Thus it is not a bad analogy to compare turbulent vibrated bed with
fluid turbulence. If this analogy is acceptable,
we may be able to get new insight about soft-to-hard phase transition.
Compared with the thermal convection,
the vibrated bed of
powder has some advantages for study.
First, our powder system has a finite number of
degrees of freedom as compared with the
 infinite number in  thermal convection.
This finiteness can
 make the numerical simulation
much easier;
for example, a low performance personal computer can
generate the hard turbulent state
in the vibrated bed of powder.
Second,
observations in real experiments of powders are expected
to be performed
much more easily than those in fluid experiments.
The displacement vectors in powder
can be  measured
photographically,
while the temperature or
the velocity in the thermal convection
can be measured
only at one or a few fixed points using special devices.
Clement and Rajchenbach\cite{Clement}
have already observed PDFs of
displacement vectors in a fluidized region and
concluded that PDFs are close to Gaussian.
However, their observation is  limited to
the center of the distribution, and we believe that much better statistics
are needed to observe
the deviation from
the Gaussian,
since the deviation can be
seen only when
the long tails of the PDF
are measured.
In fact, recent experiments of vibrated bed have
started to detect  deviations from
Gaussian PDF\cite{Warr}.

  By joining these two analogies into one, it may be interesting
to regard the soft turbulent phase
(or the solid phase)  as being composed of an
infinite cluster. An understanding of
the Gaussian velocity distribution
in this phase follows from the discussion above.
Also, the hard-turbulent phase (or the fluidized phase)
may be regarded as the state that is composed of finite clusters.
Therefore, it seems reasonable to view the soft-to-hard turbulence
transition as a percolation-like phase transition.
To put this view on a firmer foundation an intensive analysis
of the processes of clusterings
and momenta transports is required as a future task.

    In conclusion we have found both
theoretically and numerically that the velocity
distribution in a vibrated bed of powder follows
a Gaussian in the solid phase and it follows
a t-distribution with power law tails of exponent -3
in the fluidized phase.
Direct experimental confirmations of these results may be very promising.
Also, a new theoretical approach to the soft-to-hard turbulence
transition in thermal fluid convection may now be possible.

Y-h. T. thanks
Hosokawa Powder Technology Foundation and
Foundation for Promotion of Industrial Science
for the financial support, and Linux developers.
H. T. thanks H. Hayakawa and M. Takayasu for helpful discussions.

\clearpage
\begin{figure}
\begin{center}
%\input{/home/tag/doc/9502/fig.powder/fig1.inc}
% GNUPLOT: LaTeX picture
\setlength{\unitlength}{0.240900pt}
\ifx\plotpoint\undefined\newsavebox{\plotpoint}\fi
\sbox{\plotpoint}{\rule[-0.500pt]{1.000pt}{1.000pt}}%

\caption{Probability distribution functions (PDFs) of the horizontal component
$\tilde{v}_x$ of the relative velocity $\v_i^{(n)}$.
$+$: Fluidized region near surface
(the region higher than the bottom
by $10.0$.)
$\Diamond$: Solid region below the fluidized region.
Gaussian PDF is shown by the solid line
for  comparison. The abscissa is normalized by the standard deviation}
\label{fig:PDF_1}
\vspace{1cm}
%\input{/home/tag/doc/9502/fig.powder/fig2.inc}
%\input{/home/tag/doc/9504/fig.powder/fig2p.inc}
% GNUPLOT: LaTeX picture
\setlength{\unitlength}{0.240900pt}
\ifx\plotpoint\undefined\newsavebox{\plotpoint}\fi
\sbox{\plotpoint}{\rule[-0.500pt]{1.000pt}{1.000pt}}%

\caption{The same probability distribution functions (PDF)
 as Fig.\protect\ref{fig:PDF_1}, but
for the fully fluidized bed (semi-log plot and log-log plot).
The solid line shows the theoretical PDF
 of  Eq.(\protect\ref{eq:P(V).result}) with $a=0.56$.}
\label{fig:PDF_2}
\end{center}
\end{figure}

\begin{references}
\bibitem{norm}
One should notice that we scale $P(V)$ properly
because it has diverging variance.
\end{references}
\end{document}